\documentclass[iop]{emulateapj}

\def\gtsima{$\; \buildrel > \over \sim \;$}
\def\ltsima{$\; \buildrel < \over \sim \;$}
\def\gsim{\lower.5ex\hbox{\gtsima}}
\def\lsim{\lower.5ex\hbox{\ltsima}}
\shorttitle{Reverberation lag in BHXRBs}
\shortauthors{De Marco et al.}

\begin{document}

%% LaTeX will automatically break titles if they run longer than
%% one line. However, you may use \\ to force a line break if
%% you desire.

\title{Tracing the reverberation lag in the hard state of black hole X-ray binaries}

%% Use \author, \affil, and the \and command to format
%% author and affiliation information.
%% Note that \email has replaced the old \authoremail command
%% from AASTeX v4.0. You can use \email to mark an email address
%% anywhere in the paper, not just in the front matter.
%% As in the title, use \\ to force line breaks.

%% \author{B. De Marco}
%% \affil{Max-Planck-Institut f\"{u}r extraterrestrische Physik, Giessenbachstrasse 1, D-85748 Garching bei M\"{u}nchen, Germany}
%% \email{bdemarco@mpe.mpg.de}

%% \author{G. Ponti}
%% \affil{Max-Planck-Institut f\"{u}r extraterrestrische Physik, Giessenbachstrasse 1, D-85748 Garching bei M\"{u}nchen, Germany}

%% \author{T. Mu\~noz-Darias\altaffilmark{1,2}}
%% \affil{Instituto de Astrof\'isica de Canarias, 38205 La Laguna, Tenerife, Spain}

%% \and

%% \author{K. Nandra}
%% \affil{Max-Planck-Institut f\"{u}r extraterrestrische Physik, Giessenbachstrasse 1, D-85748 Garching bei M\"{u}nchen, Germany}

%% \altaffiltext{1}{Departamento de astrof\'isica, Univ. de La Laguna, E-38206 La Laguna, Tenerife, Spain}
%% \altaffiltext{2}{Department of Physics, Astrophysics, University of Oxford, Keble Road, Oxford, OX1 3RH, United Kingdom}

%% Notice that each of these authors has alternate affiliations, which
%% are identified by the \altaffilmark after each name.  Specify alternate
%% affiliation information with \altaffiltext, with one command per each
%% affiliation.

\author{B. De Marco\altaffilmark{1}, G. Ponti\altaffilmark{1}, T. Mu\~noz-Darias\altaffilmark{2, 3, 4}, K. Nandra\altaffilmark{1}} 
\altaffiltext{1}{Max-Planck-Institut f\"{u}r extraterrestrische Physik, Giessenbachstrasse 1, D-85748 Garching bei M\"{u}nchen, Germany}
\altaffiltext{2}{Instituto de Astrof\'isica de Canarias, 38205 La Laguna, Tenerife, Spain}
\altaffiltext{3}{Departamento de astrof\'isica, Univ. de La Laguna, E-38206 La Laguna, Tenerife, Spain}
\altaffiltext{4}{Department of Physics, Astrophysics, University of Oxford, Keble Road, Oxford, OX1 3RH, United Kingdom}

%% Mark off your abstract in the ``abstract'' environment. In the manuscript
%% style, abstract will output a Received/Accepted line after the
%% title and affiliation information. No date will appear since the author
%% does not have this information. The dates will be filled in by the
%% editorial office after submission.

\begin{abstract}
We report results obtained from a systematic analysis of X-ray lags in a sample of black hole X-ray binaries, with the aim of assessing the presence of reverberation lags and studying their evolution during outburst. 
We used XMM-Newton and simultaneous RXTE observations to obtain broad-band energy coverage of both the disc and the hard X-ray Comptonization components. 
In most cases the detection of reverberation lags is hampered by low levels of variability signal-to-noise ratio (e.g. typically when the source is in a soft state) and/or short exposure times. The most detailed study was possible for GX 339-4 in the hard state, which allowed us to characterize the evolution of X-ray lags as a function of luminosity in a single source. Over all the sampled frequencies ($\sim0.05-9$ Hz) we observe the hard lags intrinsic to the power law component, already well-known from previous RXTE studies.
The XMM-Newton soft X-ray response allows us to detail the disc variability.
At low-frequencies (long time scales) the disc component always leads the power law component. On the other hand, a soft reverberation lag (ascribable to thermal reprocessing) is always detected at high-frequencies (short time scales). The intrinsic amplitude of the reverberation lag decreases as the source luminosity and the disc-fraction increase. This suggests that the distance between the X-ray source and the region of the optically-thick disc where reprocessing occurs, gradually decreases as GX 339-4 rises in luminosity through the hard state, possibly as a consequence of reduced disc truncation.
\end{abstract}

%% Keywords should appear after the \end{abstract} command. The uncommented
%% example has been keyed in ApJ style. See the instructions to authors
%% for the journal to which you are submitting your paper to determine
%% what keyword punctuation is appropriate.


%% From the front matter, we move on to the body of the paper.
%% In the first two sections, notice the use of the natbib \citep
%% and \citet commands to identify citations.  The citations are
%% tied to the reference list via symbolic KEYs. The KEY corresponds
%% to the KEY in the \bibitem in the reference list below. We have
%% chosen the first three characters of the first author's name plus
%% the last two numeral of the year of publication as our KEY for
%% each reference.

%% Authors who wish to have the most important objects in their paper
%% linked in the electronic edition to a data center may do so by tagging
%% their objects with \objectname{} or \object{}.  Each macro takes the
%% object name as its required argument. The optional, square-bracket 
%% argument should be used in cases where the data center identification
%% differs from what is to be printed in the paper.  The text appearing 
%% in curly braces is what will appear in print in the published paper. 
%% If the object name is recognized by the data centers, it will be linked
%% in the electronic edition to the object data available at the data centers  
%%
%% Note that for sources with brackets in their names, e.g. [WEG2004] 14h-090,
%% the brackets must be escaped with backslashes when used in the first
%% square-bracket argument, for instance, \object[\[WEG2004\] 14h-090]{90}).
%%  Otherwise, LaTeX will issue an error. 

\section{Introduction}
\label{intro}
Luminous accreting black hole (BH) systems are powered by the same mechanism of gravitational energy release from infalling matter, most likely in the form of an accretion disc (e.g. Shakura \& Sunyaev 1973; Frank, King, \& Raine 1992). Regardless of the mass of the central black hole, $M_{BH}$, Comptonized emission from an optically thin hot plasma  (corona) of unknown geometry is also observed. This makes the primary hard X-ray continuum (e.g. Haardt \& Maraschi 1991).

In most systems, the X-ray emission from accreting BHs shows strong, aperiodic variability on a wide range of time scales. However, characteristic time scales of X-ray variability are observed which depend on the BH mass (McHardy et al. 2006; K\"{o}rding et al. 2007; Ponti et al. 2012a), matching the expected linear scaling of the size of the system with $M_{BH}$.\\
The primary X-ray continuum is reprocessed in any surrounding matter (Guilbert \& Rees 1989) including the accretion disc. If a certain degree of coherence (Vaughan \& Nowak 1997) is conserved in the reprocessing, and this does not occur exclusively in the line of sight to the central regions, light travel time delays are expected between the primary and reprocessed emission (Blandford \& McKee 1982). 
These so-called X-ray \emph{reverberation} lags (e.g. Reynolds et al. 1999, Cackett et al. 2013, Uttley et al. 2014) are a powerful diagnostics of the geometry of the corona and of the inner flow. They allow us to determine the causal relationship between the hard X-ray primary continuum and the reprocessed emission from the inner accretion disc (e.g. reflection and/or thermal reprocessing), thus probing the central region of the system.

X-ray reverberation lags have now been commonly detected in radio quiet active galactic nuclei (AGN, e.g. Fabian et al. 2009; Zoghbi et al. 2010; Emmanoulopolous, McHardy \& Papadakis 2011; De Marco et al. 2011, 2013; Kara et al. 2013a, 2013b). They have typical amplitudes of tens-to-hundreds of seconds, and appear at relatively high frequencies ($\gsim10^{-4}-10^{-3}$ Hz). These lags scale approximately linearly with the BH mass, and their amplitude and characteristic frequency are consistent with an origin within a few gravitational radii from the source of primary hard X-ray continuum emission (De Marco et al. 2013). 
These properties all suggest the lags to be the signature of reprocessing in the innermost accretion flow.
If this is indeed the case, reverberation lags would also be expected in smaller systems, where the accretor is a stellar mass BH in a binary system (BHXRB).\\
However, the light crossing time of one gravitational radius ($r_{g}/c=GM/c^{3}$) of the source is smaller in BHXRBs than in AGN by a factor determined by the difference in mass ($\gsim10^{5}$). As a consequence, even though BHXRBs can be much brighter than AGN, the number of collected photons on the $r_{g}$-crossing time is significantly smaller. This has made the detection of reverberation lags in BHXRBs more difficult.

BHXRBs are known to occasionally undergo X-ray outbursts, during which the source passes through different accretion \emph{states} and X-ray variability regimes (e.g. Miyamoto et al. 1992; Belloni et al. 2005; Belloni et al. 2011). 
Though the physical mechanism driving outbursts and state transitions is not yet fully understood, changes in the geometry of the accretion flow are thought to play a major role (Fender, Belloni \& Gallo 2004; Done, Gierli\`nski \& Kubota 2007 and references therein). 
In fact, the emission from the optically thick accretion disc appears to originate at large radii in quiescent/low luminosity states (e.g. McClintock, Horne \& Remillard 1995, Esin et al. 2001), while there is evidence of the disc reaching the innermost stable circular orbit (ISCO) during disc-dominated high-soft states (e.g. Gierli\`nski \& Done 2004; Steiner et al. 2010; Plant et al. 2014).\\
The changes in the geometry and/or physical state of the inner regions during hard and intermediate states are less clear (e.g. Done, Gierli\`nski \& Kubota 2007). 
By analogy with AGN and assuming a standard accretion disc model (Shakura \& Sunyaev 1973), the X-ray source is expected to be located at a few $r_{g}$ from the BH, and the reverberation lags should map this distance. However, if reprocessing occurs at larger radii -- e.g.  if the disc is truncated at $r_{in} >$6 $r_{g}$ -- the light travel time sampled by the reverberation lag is expected to be longer. This condition, combined with the large number of variability cycles within a single light curve, can significantly increase the possibility of detecting reverberation lags in BHXRBs.

BHXRBs have long been known to show \emph{hard X-ray lags}, in that high-energy X-ray photons are delayed with respect to those of lower energy (e.g Page, Bennetts, \& Ricketts 1981; Miyamoto et al. 1988, Nowak et al. 1999, Nowak, Wilms \& Dove 1999, Grinberg et al. 2014).
These studies of X-ray lags were restricted, for many years, to energies $\gsim 2$ keV (i.e. those covered by the \emph{Rossi X-ray Timing Explorer}, RXTE), where the spectrum is dominated by the Comptonized emission, thus suggesting the lags to be intrinsic to the power law component (e.g. Kotov, Churazov \& Gilfanov 2001).
In this spectral band the main properties of hard X-ray lags are: a characteristic decreasing trend of lag amplitude with decreasing time-scale of X-ray variability (or equivalently, with increasing frequency); an approximately log-linear dependence on energy, with the amplitude of the lags increasing as a function of energy-bands separation (e.g. Miyamoto et al. 1988, Nowak et al. 1999). 
In recent years, the use of XMM-Newton data has paved the way to studies of X-ray lags down to $\sim$0.4 keV.
In this spectral band, the disc component can be directly observed even in the power law-dominated hard state (e.g. Tomsick et al. 2008, Kolehmainen, Done, \& Di\'az-Trigo 2014), and its causal relationship with the other spectral components can be directly explored.
From the analysis of a set of XMM-Newton data of the BHXRB GX 339-4 in the hard state, Uttley et al. (2011) found new behaviour in the X-ray lags properties by extending their study into the soft X-ray band. In particular, at $E\lsim$2 keV, they observed a steepening of the log-linear dependence on energy (they detected this feature also in one dataset of Cyg X-1 and Swift J1753.5-0127), and interpreted this result as a signature of the disc variability leading the power law variations on time scales of the order of seconds or longer. Even more drastic changes are observed on variability time-scales shorter than seconds. Indeed, at such high frequencies they found a reversal of the trend, with the disc component now lagging behind the power law component by a few milliseconds. 
They interpreted these results as the first evidence of  \emph{thermal reverberation}: at high frequencies, the fraction of variable thermal disc emission produced by internal heating is low as compared to the fraction of variable disc emission due to thermalization of the power law photons in the disc (e.g. Guilbert \& Rees 1988; Wilkinson \& Uttley 2009). Within this scenario the reprocessed disc emission should track the short-time scale variations of the power law component, with a time delay equal to the light crossing time between the two emitting regions. Hence, according to this interpretation, we expect these same features to be present also during other observations, with possible deviations only related to changes of disc-corona geometry and/or physical condition throughout the outburst.

In this paper we present results obtained from a systematic search for reverberation lags in a sample of BHXRBs, including: GX 339-4, 4U 1630-47, 4U 1957+115, Cyg X-1, GRO J1655-40, GRS 1758-258, H 1743-322, Swift J1753.5-0127, XTE J1650-500, XTE J1817-330, with the aim of studying their evolution as a function of source state. We first inspected all the archived XMM-Newton observations of these sources as of August 2014. However, it turned out that, among them, only GX 339-4 has more than two archived observations with high variability power signal-to-noise (S/N) up to sufficiently high frequencies (i.e. $\sim$10 Hz). This is a necessary condition for the detection and monitoring of reverberation lags. 
For this reason we will first focus on the results obtained from the detailed analysis of the XMM-Newton and simultaneous RXTE observations of GX 339-4 (Section \ref{sec:gx}). Results obtained from the analysis of the rest of the sample will be reported in Section \ref{sect:other}.

\section{The case of GX 339-4}
\label{sec:gx}

\subsection{Data reduction and analysis}
\label{sec:data}
We analyzed the publicly available XMM-Newton observations of GX 339-4 as of August 2014. Among the observations analysed, a total of four (all carried out during the hard state) turned out to have sufficiently high variability power S/N ratio to enable timing analysis. 
The other observations (mostly carried out during intermediate or soft states of the source) are either too short or have too low intrinsic variability power to allow us to obtain reliable spectral-timing measurements (see Table \ref{tab:non-det}).
Note that the selection and data reduction criteria are the same as used in De Marco et al. (2015, hereafter DM15), thus we refer to this paper (specifically section 2 and table 1) for a more detailed description.

The four XMM-Newton observations of GX 339-4 presented in this paper belong to two different outbursts, and were carried out on 16-18 March 2004, 26 March 2009, and 28 March 2010 (with corresponding observation IDs: 0204730201, 0204730301, 0605610201, 0654130401). Since the first two belong to consecutive orbits, we combined them to obtain better statistics. We extended the spectral band coverage up to $\sim$30 keV, by also analysing seven simultaneous RXTE observations (IDs: 90118-01-05-00, 90118-01-06-00, 90118-01-07-00, 94405-01-03-00, 94405-01-03-01, 94405-01-03-02, 95409-01-12-01).

For XMM-Newton, we used only EPIC-pn (Str\"uder et al. 2001) data taken in timing-mode, and for RXTE either Good Xenon or Generic Event Proportional Counter Array (PCA; Jahoda et al. 2006) data. For the PCA we considered data from all the layers of the Proportional Counter Units (PCU) simultaneously and continuously switched on during the single observations. Data reduction has been conducted following standard procedures and using the latest XMM-Newton calibration files (CCF) relevant for the Timing-mode (as of August 2014). The net exposure time for the XMM-Newton observations is of $\sim$175 ks (2004 observations), $\sim$32 ks (2009 observation), and $\sim$33 ks (2010 observation). The net exposure time for the RXTE observations is of $\sim$10 ks (2004 observations), $\sim$8 ks (2009 observations), and $\sim$19 ks (2010 observations).

During the selected observations, the source is in a hard state, with Eddington-scaled luminosities ($E=3-30$ keV) of $L/L_{Edd}\sim 0.007$ (2009 observations), $\sim 0.02$ (2004 observations), and $\sim 0.07$ (2010 observations)\footnote{Here we assume a distance of 8 kpc and a BH mass of 8 M$_{\odot}$. These values correspond to average values derived from the currently available estimates and uncertainties on these two parameters (Hynes et al. 2003, 2004, Mu\~{n}oz-Darias et al. 2008).}, and almost constant hardness ratios (defined as the ratio between the 13-30 keV to the 3-13 keV RXTE flux, Dunn et al. 2010), spanning the range $\sim$0.8--1 (see also figure 1 of DM15). Hereafter we adopt the same nomenclature as used in DM15, and refer to the analyzed observations as low-, medium-, and high-luminosity. 

For the timing analysis we adopted the same sampling used to compute the power spectra in DM15: for each observation and energy band we computed the cross-spectra from light curve segments of 21 s, 59 s, and 500 s length (respectively for the high-, medium-, and low-luminosity observation), and averaged them to obtain estimates of the frequency-dependent cross-spectrum. 
This choice of light curve segments is dictated by the necessity of excluding gaps due to telemetry drop-outs (which occur mostly when the target source is very bright) in XMM-Newton data.
The light curves have been extracted with a time resolution of 10 ms, thus ensuring a wide frequency coverage (up to about four decades for the longest segments choice). Nonetheless, we limited the analysis to frequencies $\nu\lsim$10 Hz, since above this threshold the effects of channel cross-talk induced by Poisson fluctuations are not negligible and significantly affect the lag estimate (Lewin, van Paradijs \& van der Klis 1988; Vaughan et al. 1999).

\begin{figure*}
\includegraphics[width=\textwidth]{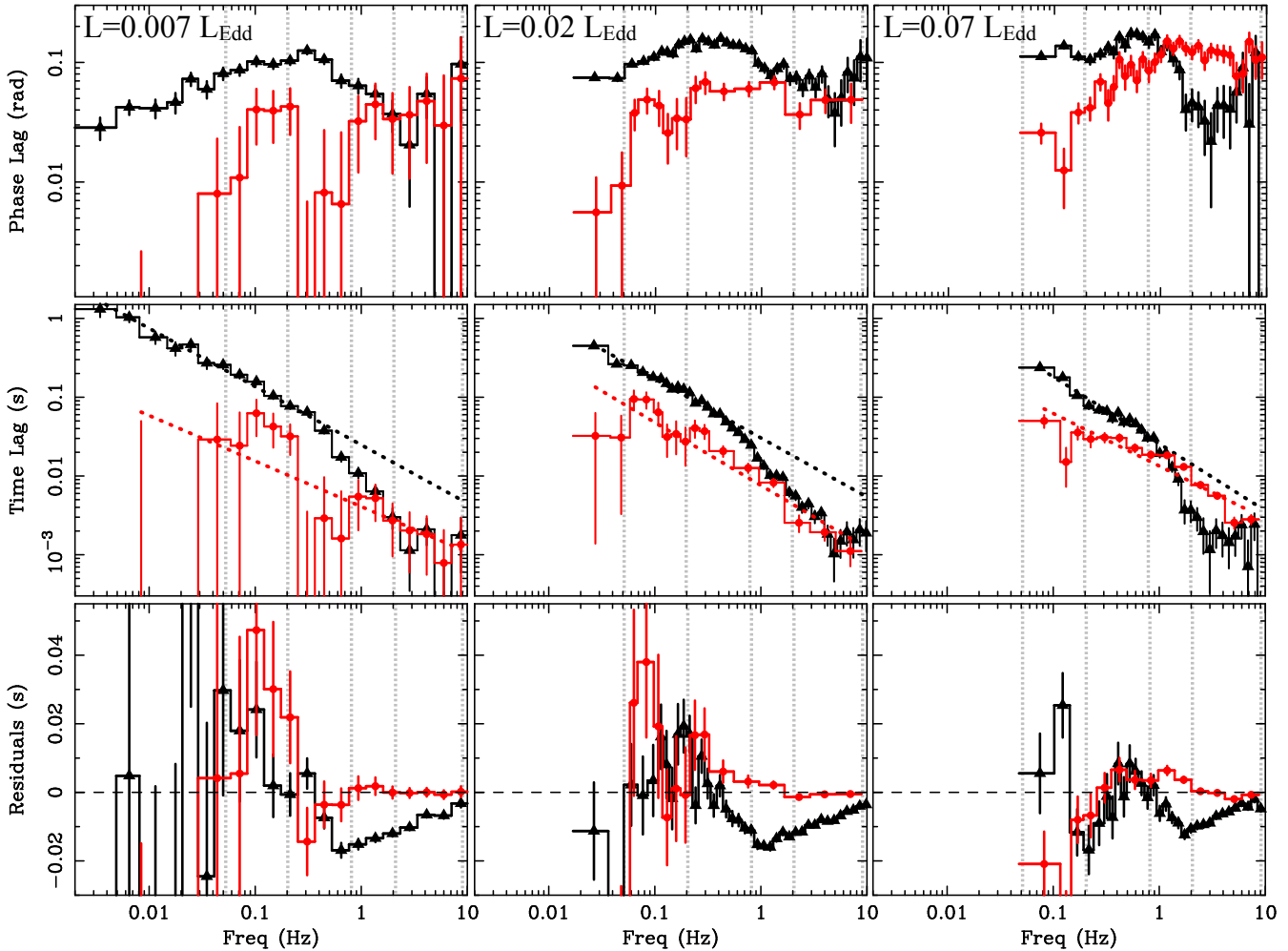}
\caption{Phase and time lags of GX 339-4 in the Fourier-frequency domain. Each column refers to each of the analysed observations, with 3-30 keV Eddington scaled luminosities: 0.007$L_{Edd}$, 0.02$L_{Edd}$, 0.07$L_{Edd}$. The black and red data points/curves in each plot correspond respectively to the phase/time lags between the \emph{soft} and \emph{hard} energy bands (using XMM-Newton data), and between the \emph{hard} and \emph{very hard} energy bands (using RXTE data).
The rows display, from top to bottom: the phase lags as a function of frequency; the corresponding time lags as a function of frequency, and their  best-fit power law model (note that for the \emph{soft} vs. \emph{hard} data, the model refers to the best-fit at low-frequencies, $\nu<0.6$ Hz, only, see Sect. \ref{sect:lag-freq}); the time lag residuals to the best-fit power law model. The vertical dotted lines (grey) mark the frequency intervals used for the extraction of the lag-energy spectra in Sect. \ref{sect:lag-energy}.}
\label{fig:coherence}
\end{figure*}

\subsection{Lags in Fourier-frequency domain}
\label{sect:lag-freq}
We first analysed the X-ray lags as a function of Fourier-frequency, following standard techniques (e.g. Nowak et al. 1999, Uttley et al. 2014).
The cross-spectrum has been computed between the same energy bands adopted in DM15, i.e. 0.5-1.5 keV and 2-9 keV (using XMM-Newton data), and between bands 2-9 keV and 10-30 keV (using RXTE data). Using the same nomenclature as in DM15, hereafter we will refer to the three energy bands as \emph{soft}, \emph{hard}, and \emph{very hard}.\\
Phase ($\phi$) and time lags ($\tau=\phi/2\pi\nu$, where $\nu$ represents the centroid of each frequency bin) as a function of Fourier-frequency for each of the observations analysed are shown in Fig.~\ref{fig:coherence}. Here we used a standard logarithmic frequency-rebinning.

The phase/time lags are characterized by positive amplitudes over all the sampled frequencies, meaning that light curves in harder bands are delayed with respect to light curves in softer bands. According to past studies, a power law model for the time lags, with index $\sim -0.7, -0.8$, has been proved to qualitatively describe the underlying decreasing trend as a function of frequency in BHXRBs (e.g. Nowak et al. 1999; Pottschmidt et al. 2000).

This is indeed the case for the \emph{hard} vs. \emph{very hard} time lags, already well known thanks to the extensive RXTE monitoring campaigns. Fig.~\ref{fig:coherence} clearly shows that complex structures are superimposed on this underlying power law trend. These structures are commonly observed in BHXRBs (e.g. Miyamoto et al. 1992; Nowak et al. 1999). As is more easily visible in the phase lag-frequency plots (top panels of Fig.~\ref{fig:coherence}), the complexity of the lag profiles increases when the \emph{soft} X-ray band is considered (in agreement with Grinberg et al. 2014). 
In particular, phase lags between the \emph{soft} and the \emph{hard} band always show a maximum, followed by a drop at high-frequencies. 
The position of the maximum is shifted to higher frequencies as the luminosity increases, and its amplitude also increases.

To better characterize the deviations of the time lag profile from a smooth decreasing trend as a function of frequency, we carry out a fit with a simple power law model.
For the \emph{hard} vs. \emph{very hard} lag, independently of luminosity, we derive a power law slope consistent with the values commonly found in the literature (e.g. Nowak  et al. 1999) from the analysis of RXTE light curves of BHXRBs. Specifically we estimate slopes of $-0.58\pm0.25$, $-0.80\pm0.05$, and $-0.67\pm0.04$, respectively for the low-, medium-, and high-luminosity observation (these models are over plotted on the red data points in Fig.~\ref{fig:coherence}, middle panels).
On the other hand, the \emph{soft} vs. \emph{hard} lags show significantly more structured profiles, substantially deviating from a power law. The fits also return much steeper trends, with best-fit slopes of $-0.93\pm 0.02$, $-0.96\pm0.01$, and $-1.12\pm0.03$  respectively for the low-, medium-, and high-luminosity observation.

As pointed out in Uttley et al. (2011), the steepening of the \emph{soft} vs. \emph{hard} lag is due to a drop in lag amplitude at high frequencies, which we observe in all the observations analysed. 
Nonetheless, the low-frequency ($\lsim0.6$ Hz) \emph{soft} vs. \emph{hard} lags appear to exhibit a similar trend as observed in the \emph{hard} vs. \emph{very hard} lags over the entire frequency range analysed. We verify this, by excluding the high frequencies ($\gsim0.6$ Hz) in the \emph{soft} vs. \emph{hard} lags and performing the fits again. We obtain best-fit slope values of $-0.74\pm0.04$, $-0.75\pm0.02$, and $-0.85\pm0.06$ (these models are over plotted on the black data points in Fig.~\ref{fig:coherence}, middle panels), respectively for the low-, medium-, and high-luminosity observation (residuals to the power law models are plotted in Fig.~\ref{fig:coherence}, bottom panels). These values are consistent with those obtained at higher energies, and point to the same underlying variability process producing the low-frequency hard-lags in softer X-ray bands.

The high-frequency drop of time lag amplitude (producing the negative residuals below the low-frequency power law trend shown in bottom panels of Fig.~\ref{fig:coherence}, black data points) suggests that the hard lags are suppressed above a certain frequency. This behaviour is observed only when the \emph{soft} X-ray band is considered, which contains significant contribution from the disc black-body emission (see Sect. \ref{sect:rev}), thus suggesting a link between the two.

Reverberation should produce small-amplitude, negative lags between the power law and the disc emission (i.e. with the disc photons lagging behind the power law photons). This negative-lag component would combine with the overall hard lags at all frequencies, thus reducing their observed amplitude. However, even if a reverberation component is present over a broad range of frequencies, it will be detectable only at high frequencies, where the amplitude of the hard lag is sufficiently small, e.g. as a consequence of high-frequency suppression. 

In AGN reverberation lags are observed at relatively high frequencies (of the order of $\gsim10^{-3}$ Hz for relatively low mass sources of $\sim10^{6}$ M$_{\odot}$).  
As in BHXRBs, the low-frequencies are, instead, dominated by a hard lag component, well-described by a power law with slope close to $-1$ (e.g. Papadakis, Nandra \& Kazanas 2001). 
The high-frequency drop of the hard lag component enables detection of the reverberation lag (whose amplitude is of the order of tens of seconds for a $\sim10^{6}$ M$_{\odot}$ source). 
By approximately rescaling the amplitude and the frequency of the reverberation lag observed in AGN for the difference of mass with GX 339-4 (specifically assuming a factor $\sim10^{5}$), reprocessing from the inner regions of the accretion disc (thus producing a time delay of the order of $\sim10^{-4}$ s) would dominate at frequencies of the order of $\gsim100$ Hz in this BHXRB, well outside the frequency window we are considering (i.e. $<$10 Hz).
However, we expect the amplitude of the reverberation lag to scale linearly with the distance from the disc region where reprocessing occurs. If, for example, this region is not at the ISCO during the hard state (see Sect \ref{intro}), e.g. assuming a factor $\sim$10 larger distance, 
a higher-amplitude reverberation lag, of the order of $\sim10^{-3}$ s, is expected. In this case, given the observed \emph{soft} vs. \emph{hard} lag-frequency spectra of GX 339-4 (Fig.~\ref{fig:coherence}, middle panels), a significant contribution from the reverberation lag should be observable at frequencies $\gsim$1 Hz, though a drop to negative values (as seen in AGN) would probably appear at higher frequencies.\\
If the observed high-frequency drop in the lag-frequency spectra of GX 339-4 is associated with the emergence of a reverberation lag, we expect to observe its signature in the frequency-resolved, energy-spectra of the time lags, at the energies where the disc black body emission significantly contributes to the total flux.

\begin{figure*}[!h]
\includegraphics[width=\textwidth]{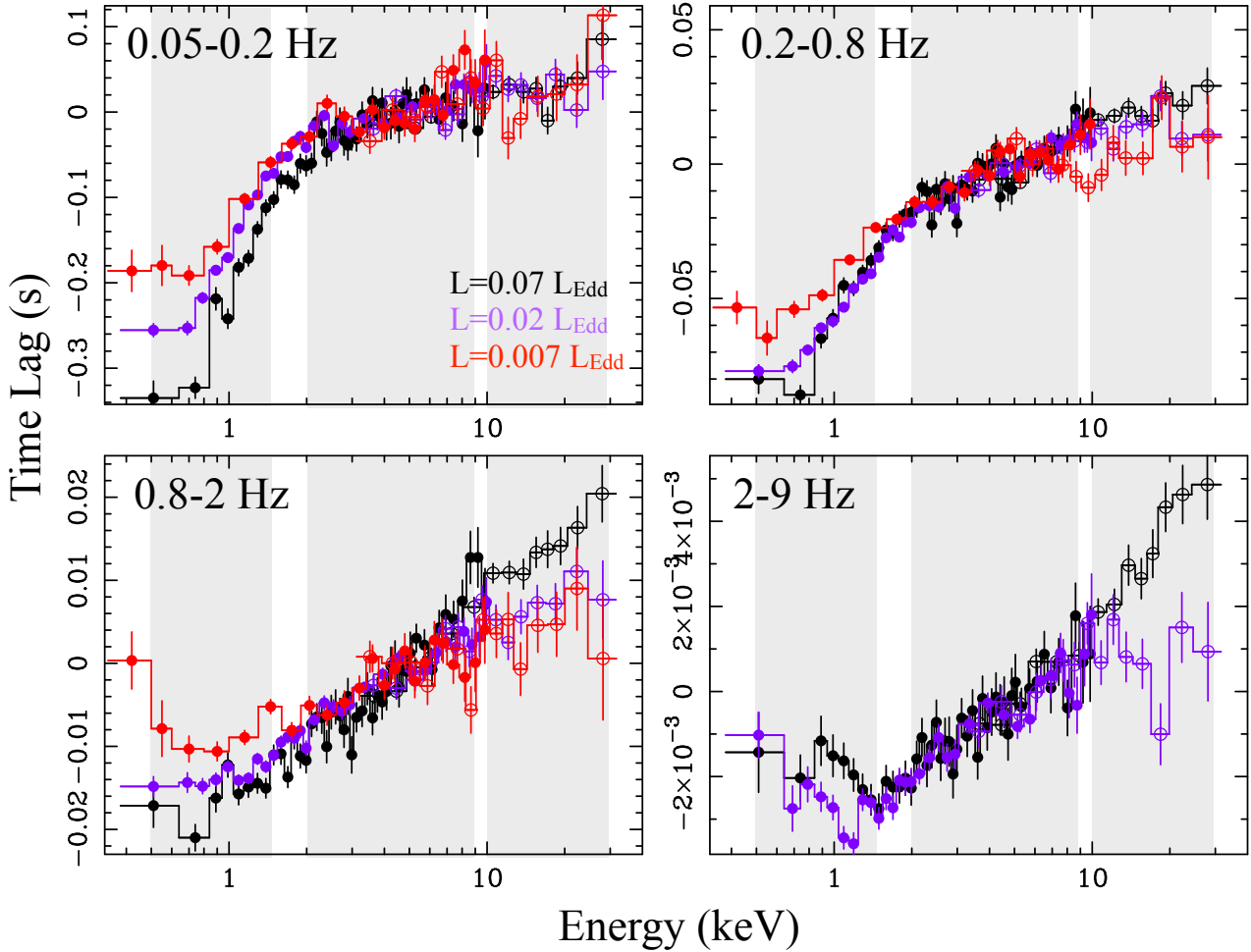}
\caption{Time lag-energy spectra of the hard state observations of GX 339-4. The spectra are computed in four frequency intervals. In each plot the spectra from the different observations (in different colors) are over plotted for comparison. XMM-Newton (filled symbols) and RXTE (open symbols) lag-energy spectra are computed separately, rescaled to the same reference, and plotted in the same color for each observation. The shaded areas mark the \emph{soft}, \emph{hard}, and \emph{very hard} bands used for the study of lags as a function of frequency (Sect. \ref{sect:lag-freq} and Fig.~\ref{fig:coherence}).}
\label{fig:lagE}
\end{figure*}

\begin{figure*}[!h]
\centering
\includegraphics[width=\textwidth]{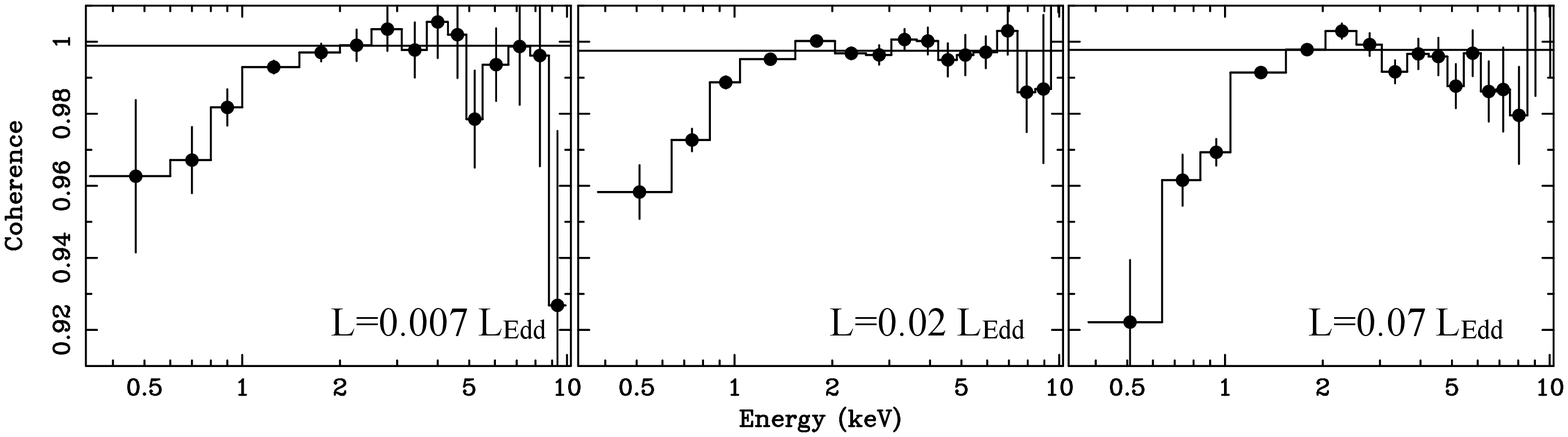} 
\caption{The energy-dependence of the intrinsic coherence of GX 339-4 as measured in the 0.05-0.2 Hz frequency interval. The horizontal line is the best-fit constant model to the high-energy (E$>$2 keV) coherence. Note that, despite the small drop at low energies, the coherence is high ($>0.9$) at all energies.}
\label{fig:coher_lowf}
\end{figure*}

\begin{table}
\caption{Log-linear fits of the lag-energy spectra}  
\label{tab:fits}
\centering
\begin{tabular}{c c c}

\tableline
	     &    $\alpha_{1}$        & $\alpha_{2}$  \\
\tableline
 & \multicolumn{2}{ c }{0.05-0.2 Hz} \\
\tableline
High-L    &   0.441$\pm$0.044 &  0.064$\pm$0.014   \\  
Medium-L    &   0.381$\pm$0.024 &  0.078$\pm$0.013  \\ 
Low-L    &   0.323$\pm$0.050 &  0.069$\pm$0.025   \\ 
\tableline
 & \multicolumn{2}{ c }{0.2-0.8 Hz} \\
\tableline
High-L    &  0.111$\pm$0.012 &   0.045$\pm$0.004  \\ 
Medium-L    &  0.120$\pm$0.007 &   0.036$\pm$0.004 \\ 
Low-L    &  0.071$\pm$0.013  & 0.016$\pm$0.007 \\ 
\tableline
 & \multicolumn{2}{ c }{0.8-2 Hz} \\
\tableline
High-L    &  0.026$\pm$0.006 &   0.028$\pm$0.002  \\ 
Medium-L    &  0.025$\pm$0.003 &   0.016$\pm$0.002  \\ 
Low-L    & 0.008$\pm$0.007  & 0.009$\pm$0.003 \\ 
\tableline
 & \multicolumn{2}{ c }{2-9 Hz} \\
\tableline
High-L    &  0.005$\pm$0.002 &   0.006$\pm$0.001 \\ 
Medium-L    &  0.004$\pm$0.002 &   0.003$\pm$0.001  \\ 
Low-L    & 0.002$\pm$0.004  & 0.004$\pm$0.001 \\ 
\tableline
\end{tabular}
\tablecomments{Best-fit values of the slope parameter as obtained from the log-linear fits of the lag-energy spectra of GX 339-4 at E$\geq 1$ keV. $\alpha_{1}$ and $\alpha_{2}$ refer to fits carried out respectively in the low-energy (E$=1-2.5$ keV) and high-energy (E$\geq2.5$ keV) portion of the lag spectra.}
\end{table}

\subsection{Frequency-resolved lag-energy spectra}
\label{sect:lag-energy}
We investigated the energy-dependence of the observed time lags in more detail, by computing their frequency-resolved energy-spectra (see Uttley et al. 2014 for a detailed description of the techniques). To this aim, we computed the lags between a reference band and a series of small, adjacent energy bins. We averaged the lags within frequency intervals of interest and plot them as a function of energy. 
To obtain a broader spectral band coverage, the lag-energy spectra have been measured, separately, from both the XMM-Newton and RXTE data sets. 
To ensure high S/N ratio in the lag-energy spectra, while retaining relatively good spectral resolution, we used a wide energy interval as the reference band. However, to avoid spurious effects due to correlated Poisson noise between the reference and any given energy bin, the energy bin is removed from the reference light curve before computing the lags.  
At low energies ($\lsim 1$ keV) the resulting lag-energy spectra might be affected by effects related to incomplete charge collection close to the EPIC pn detector surface (e.g. Popp et al. 1999, 2000). These influence the detector response, so that, for monochromatic X-rays of a given photon energy, a fraction of photons will be detected at slightly lower energies (this fraction increases as the energy of the photons decreases). The result would be to mix the lags from adjacent bins, thus diluting the intrinsic lag at soft X-ray energies, if present. We 
verified that the chosen width of energy bins minimizes this effect, such that this bias yields deviations on the lag estimates smaller than the lag uncertainty.

\begin{figure*}[!t]
\centering
\includegraphics[width=\textwidth]{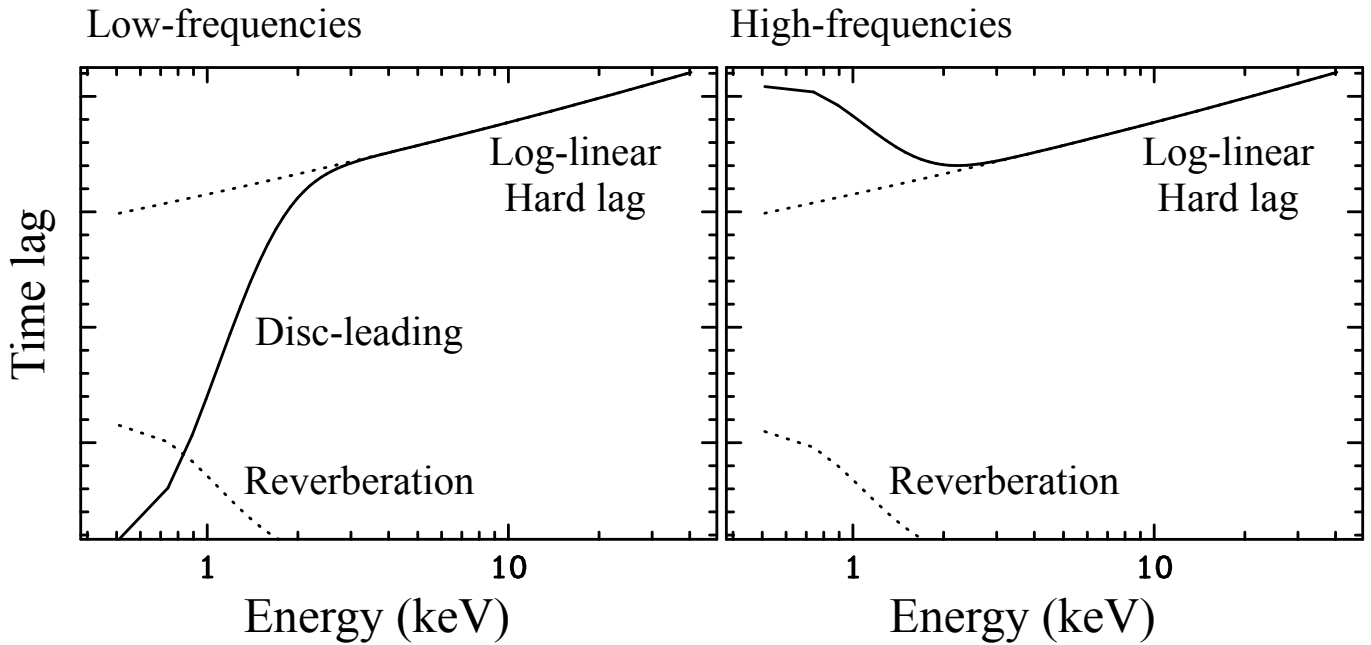}
\caption{Schematic illustration of the main features characterizing the time lag-energy spectra of GX 339-4 at low and high frequencies. Note that the observed lag as a function of energy is not a simple sum of the lags associated with the different components (see also discussion in Sect. \ref{sect:Ldepend}).}
\label{fig:scheme}
\end{figure*}

We adopted the 0.5-10 keV band as the reference for the XMM-Newton data, so as to get high quality lag spectra down to energies where the reverberation lag should be observed. On the other hand we chose the 3-10 keV band as the reference for the RXTE data. Note that phase and time lags are not affected by flux calibration issues when comparing data from different instruments, and any relative shift is due to the choice of different reference bands.
However, to allow for a direct comparison of the lag-energy spectra obtained separately from the EPIC pn and PCA detectors, we reported them to the same reference band, i.e. 3-10 keV. To this aim we phase-shifted the EPIC pn lags by an amount which depends on the relative phase between light curves in the initial (0.5-10 keV) and the new (3-10 keV) reference band\footnote{We verified this procedure by comparing the results with those that would be obtained by computing the XMM-Newton lag-energy spectra using directly the 3-10 keV band as the reference band. We confirm our procedure yields consistent results, with the advantage of retaining higher S/N lag spectra. Indeed, although the EPIC-pn response is softer than the PCA response, the average energy of variable photons (as derived from the covariance spectrum, Uttley et al. 2014) in the 3-10 keV reference band differs by only $\sim 0.6$ keV between the two detectors.}.\\
Our results are shown in Fig.~\ref{fig:lagE} (where the shaded areas mark the energy bands that were used for the lag-frequency spectra in Sect. \ref{sect:lag-freq} and Fig.~\ref{fig:coherence}) for the frequency intervals: 0.05-0.2 Hz, 0.2-0.8 Hz, 0.8-2 Hz, and 2-9 Hz (for reference these intervals have been indicated by vertical dotted lines in the plots of Fig.~\ref{fig:coherence}). These intervals are common to all the analysed data sets, and have been chosen so as to continuously sample the complex structures observed in the phase/time lags as a function of frequency (i.e. humps, dips, and plateaus). The intrinsic coherence as a function of energy (Vaughan \& Nowak 1997) has been checked. We found the coherence to be remarkably high (\gsim 0.9) at all energies and for all the frequency intervals (e.g. the coherence vs. energy for the 0.05-0.2 Hz frequency interval is shown in Fig.~\ref{fig:coher_lowf}).
The only exception is given by the 2-9 Hz interval of the low-luminosity observation, where the coherence is much noisier thus increasing the uncertainty on the lag measurements. For the sake of clarity we omit the corresponding lag-energy spectrum in Fig.~\ref{fig:lagE}, since it is mostly unconstrained. \\
All the broad band lag-energy spectra of GX 339-4 show similar, frequency-dependent, trends (see Fig.~\ref{fig:lagE}). In particular, at high energies a log-linear dependence always holds, while a deviation is observed in the soft X-ray band both at low and high frequencies. 
At low frequencies a steepening of the high-energy log-linear trend is observed at $E\lsim 2$ keV. In this same band a reversal of this trend is observed at high frequencies.
These features have been ascribed to the disc component (Uttley et al. 2011), whose intrinsic variability leads the power law variations at low-frequencies. On the other hand, at high-frequencies the disc thermally-reprocessed emission tracks the power law variations, so that a reverberation lag is observable in the soft X-ray band. According to this interpretation signatures of the disc leading the variability at low-frequencies and of disc reverberation at high-frequencies should be detected in every hard state observation, as we indeed show. We confirm the detection of these two features during the medium-luminosity observation of GX 339-4 as reported in Uttley et al. (2011). We find the same low-frequency disc leading feature reported in Uttley et al. (2011) during the low-luminosity observation (as well as in one hard-state observation of Cyg X-1 and Swift J1753.5-0127).\\ 
In addition, we show that both the low-frequency disc leading and the high-frequency disc reverberation appear to be a normal property throughout the hard state.
Moreover, we observe significant differences as a function of luminosity, that might be the consequence of changes of the disc-corona structure/geometry. To highlight these differences, in Fig.~\ref{fig:lagE} we have over plotted, for each frequency interval, the measured time lag-energy spectra for all the observations at different luminosities. 
The main features characterizing the lag-energy spectra of GX 339-4 can be summarized as follows:
\begin{itemize}
\item {\bf $\nu<$ 0.8 Hz:} at high energies (i.e. E$\gsim2$keV) all the lag-energy spectra are consistent with one another, and show small relative delays (between different energy bins) up to the highest sampled energy bands. 
On the other hand, at low energies all the spectra break to a steeper profile (corresponding to larger relative delays), but the slope tends to increase with luminosity (Table \ref{tab:fits}). In other words, the soft X-ray bands (E$\lsim$ 2keV) always lead the hard X-ray bands (E$\gsim$ 2keV), but the amplitude of the corresponding time delay increases with luminosity. At all luminosities the lag-energy spectra flatten out towards very soft energies.
\item {\bf $\nu>$ 0.8 Hz:} the lag-energy spectra can be broadly described by a single-slope, luminosity-dependent, log-linear model (with steeper slopes associated with higher luminosities Table \ref{tab:fits}). However, at soft X-ray energies disc reverberation is observable in all the spectra. The energy at which this trend-reversal is detected increases with luminosity (i.e. E$\sim$0.8 keV, 1.2 keV, and 1.5 keV, respectively for the low-, medium-, and high-luminosity observations, see Sect. \ref{sect:rev}). 
\end{itemize}
In Fig.~\ref{fig:scheme} we show a synthetic picture of the main features characterizing the lag-energy spectra of GX 339-4 at low and high frequencies, which is meant to illustrate their energy dependence. Note, however, that the observed lag as a function of energy is not the simple sum of the lags associated with the different components (as discussed in Sect. \ref{sect:Ldepend} these should combine in a more complex way).
Notably, the switch in the lag-energy dependence (from disc leading to disc lagging), as well as the high-frequency drop in the lag-frequency profiles (see Fig.~\ref{fig:coherence}), occur at approximately the same frequencies, i.e. above $\sim$0.8 Hz. In DM15 we showed that in this range the high-frequency Lorentzian dominates the power spectra of GX 339-4. Thus the detected changes in the lag-energy and lag-frequency spectra are associated with the emergence of this variability component. 
In the likely hypothesis that the high-frequency Lorentzian represents the X-ray variability power from the inner accretion flow, the detection of a reverberation lag is expected right at these frequencies.\\

\subsection{The reverberation lag}
\label{sect:rev}
\begin{figure*}[!t]
\centering
\includegraphics[width=\textwidth]{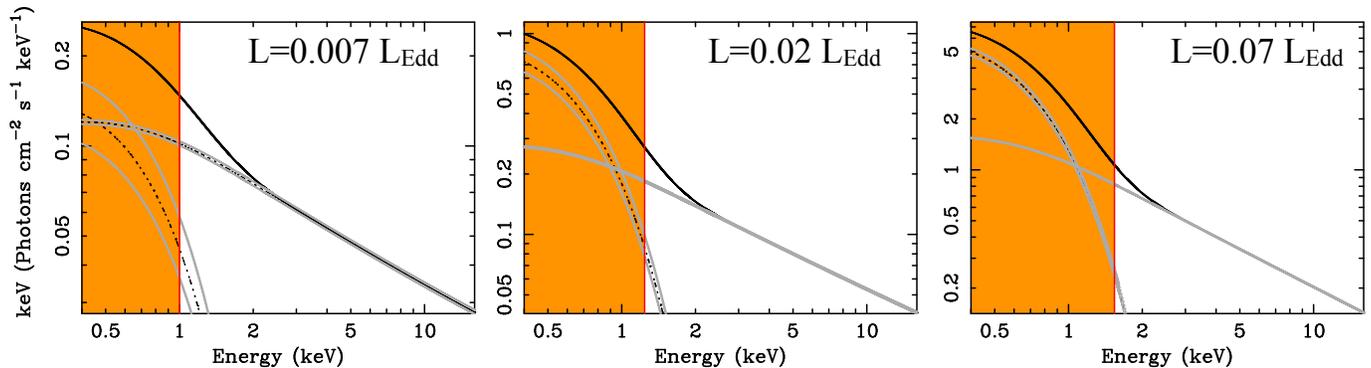}
\caption{Best-fit models for the continuum of GX 339-4 (in the XMM-Newton data sets). The model used is $tbabs*[diskbb+nthcomp]$, though in the plots the absorption component is removed. The shaded area marks the energies where the reverberation lag is observed in the high-frequency lag-energy spectra of Fig.~\ref{fig:lagE}.}
\label{fig:spectra}
\end{figure*}

\begin{figure}
\centering
\includegraphics[width=8.cm]{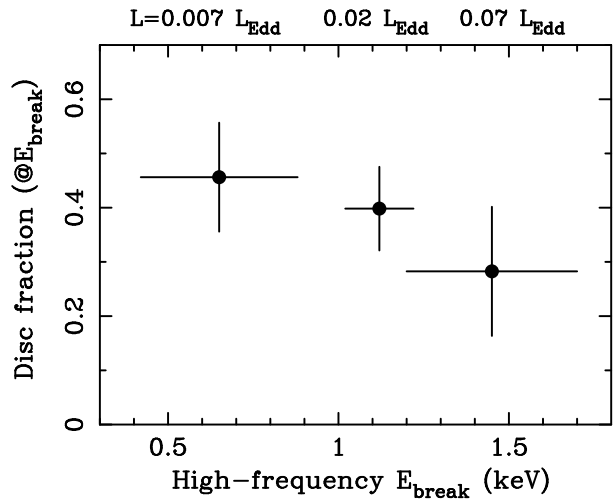} 
\caption{The energy at which a trend reversal is observed in the high-frequency lag-energy spectra (obtained fitting the spectra with a broken power law model) as compared to the disc flux fraction (i.e. the ratio between the disc and the disc-plus-power law flux) in the energy bin of the reversal.}
\label{fig:upt_disc}
\end{figure}

At high-frequencies the \emph{soft} vs. \emph{hard} time lag-frequency spectra (Fig.~\ref{fig:coherence}, middle panels) show a drop. At these same frequencies, the lag-energy spectra show that a soft band time delay starts to be observable above the log-linear hard lags trend.
This is consistent with the emergence of a reverberation lag associated with the disc thermal emission, as already observed in one observation of GX 339-4 by Uttley et al. (2011) and strengthened here by the detection of the same lag in all the analysed hard state XMM-Newton observations of GX 339-4.
To reinforce this interpretation, we fit the XMM-Newton spectra of GX 339-4 with a simple model for the continuum (i.e. $tbabs*[diskbb+nthcomp]$) with the Comptonization high energy cut-off fixed at 100 keV (as appropriate for hard states, e.g. Motta, Belloni \& Homan 2009), the seed photon temperature tied to the inner disc temperature, and excluding the range of energies dominated by the Fe K$\alpha$ line. Moreover, we exclude the energies around the edges of the response matrix at $\sim1.8$ keV and $\sim$2.2 keV, which produce strong residuals, probably an artefact of uncorrected X-ray loading (XRL) effects and/or charge transfer inefficiency (CTI) of the EPIC-pn timing mode (Kolehmainen et al. 2014). Note that we are not interested in a detailed spectral modeling, but rather in a broad characterization of the primary power law and disc component.\\
Fig.~\ref{fig:spectra} shows the best-fit model for each observation, once the galactic absorption component has been removed from the plots. Note that the best fit inner disc temperature is $kT_d\sim$0.22-0.26 keV, thus the low-energy rollover of the Comptonization component is not observable within the energy window analysed. 
In the fits we let the cold absorption column density parameter, $N_{H}$, be free. Nevertheless, we obtain best-fit values in the range $N_{H} \sim 5-6 \times 10^{21}$ cm$^{-2}$, in agreement with values reported in the literature (Dickey \& Lockman 1990; Kong et al. 2000).
We also show the 90 percent confidence contours (grey curves) obtained by varying the normalization of the power law and disc components. Overplotted in orange is the shaded area which marks the range of energies where the reverberation lag has been detected (namely from the start to the end point of the trend reversal observed below $\sim$2 keV, in the two bottom panels of Fig.~\ref{fig:lagE}).
A direct link between the emergence of a significant disc component and the presence of the lag is evident. 
Moreover, as noticed in the previous Section, the energy at which the reverberation lag starts to be observable above the broad-band log-linear trend varies as a function of luminosity. This behaviour matches well the shift, as a function of luminosity, of the energy at which the disc component starts to contribute significantly to the total X-ray flux. 
We parametrized the energy at which the trend reversal is seen in the high-frequency lag-energy spectra fitting them with a simple broken-power law model. In Fig. \ref{fig:upt_disc} we compare the estimated break energy with the disc-to-total X-ray flux ratio at this energy. We find that within the energy bin where the reverberation lag emerges the disc component always contributes $\sim$40 percent of the total (disc-plus-power law) flux.

Nonetheless, it is well-known that the \emph{hard} energy bands, spectrally dominated by the power-law component, display significant aperiodic variability over a very broad range of time scales (about four orders of magnitude, see DM15). 
If part of this emission is thermally reprocessed in the disc, then signatures of a reverberation lag are expected even at lower frequencies. This aspect is explored in Sect. \ref{sect:Ldepend}, where we analyse in more detail the lag-energy spectra of GX 339-4, to study the luminosity dependence of \emph{soft} X-ray band disc-leading and disc-reverberation lags. We show that they are characterized by opposite trends, which can be both ascribed to intrinsic variations of the amplitude of the reverberation lag.

\subsection{Luminosity dependence of soft X-ray band lags}
\label{sect:Ldepend}
We analysed the variations of the low-frequency disc-leading feature and of the high-frequency reverberation feature in the time lag-energy spectra as a function of 3-30 keV Eddington-scaled luminosity and disc-to-power law flux ratio at energies E$\lsim$3 keV (as obtained from the best fit model described in Sect. \ref{sect:rev}).

\begin{figure*}
\centering
\includegraphics[width=\textwidth]{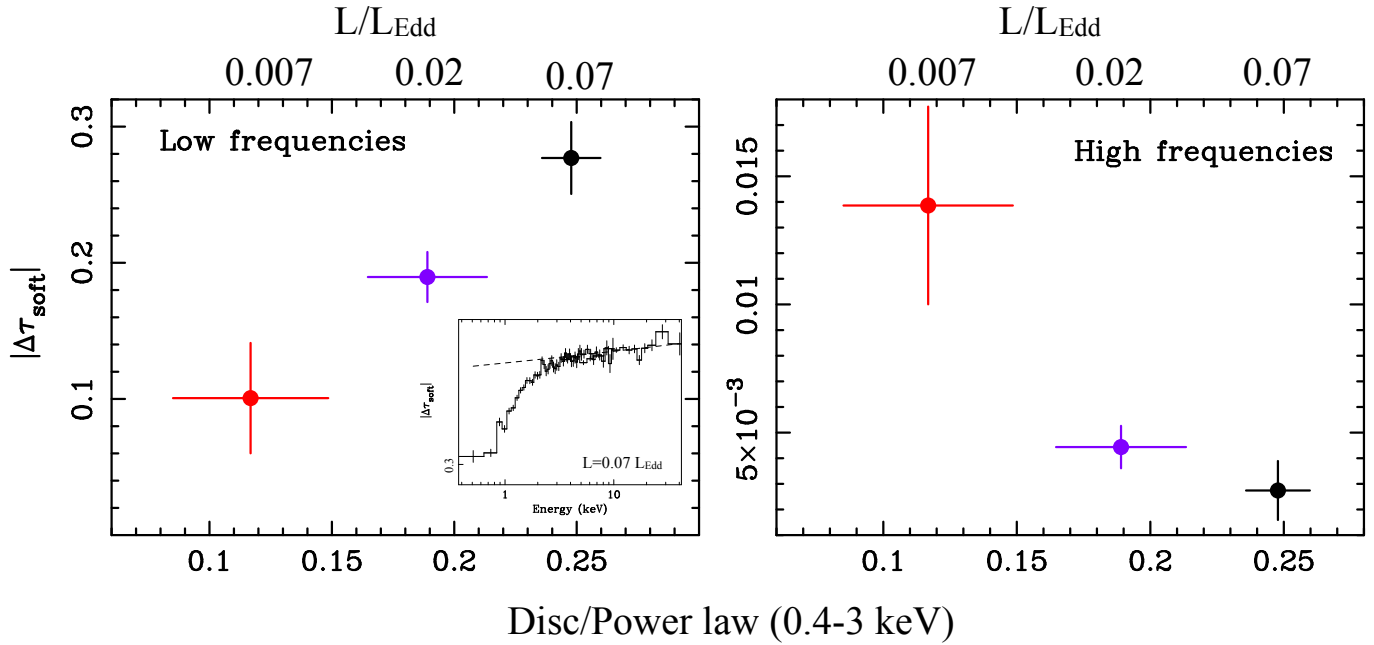} 
\caption{The amplitude of the soft-band disc-leading (left panel) and reverberation (right panel) features, respectively in the low-frequency (0.05-0.2 Hz) and high-frequency (0.8-2 Hz, red data point, and 2-9 Hz, blue and black data points) lag-energy spectrum of GX 339-4 as a function of disc-to-power law flux ratio in the 0.4-3 keV range. The inset in the left panel illustrates how the amplitude of the soft-band disc-leading feature is determined from the extrapolation in the soft band of the high-energy, log-linear best fit model (dashed lines).
The different colors refer to the observations at different luminosities (red, blue, and black respectively for the low-, medium-, and high-luminosity observations). Luminosities are also reported (in the 3-30 keV band and in units of Eddington luminosity) on the top of the plot. }
\label{fig:lowhighf}
\end{figure*}

To parametrize the amplitude of the soft-band disc-leading feature in the time lag-energy spectra we computed the maximum difference (in absolute value), $\Delta\tau_{soft}$, between the measured lag at E$<$3 keV, and the extrapolation in this band of the best-fit log-linear model for the high-energy (E$>3$ keV) hard lags (dashed line in the inset of Fig.~\ref{fig:lowhighf}, left panel). 
Results shown in Fig.~\ref{fig:lowhighf} (left panel) refer to the 0.05-0.2 Hz frequency interval (the errors on $\Delta\tau_{soft}$ are computed combining the lag measurement errors and the uncertainties on the best-fit log-linear model). We followed a similar procedure for the high-frequency lag-energy spectra (i.e. the 0.8-2 Hz interval for the low-luminosity observation and the 2-9 Hz interval for the medium- and high-luminosity observations, bottom panels of Fig.~\ref{fig:lagE}), where the soft-band reverberation lag is observed.
We parametrized the amplitude of the reverberation lag as the maximum difference (in absolute value), $\Delta\tau_{soft}$, between the measured lag at $<$2 keV and the extrapolation in the soft band of the high-energy (E$\gsim$1.3 keV) best-fit log-linear model\footnote{Note that, in the fits, we discarded the data above $\sim$10 keV, because in this band the high-frequency lag-energy spectra show signs of deviations from a single-slope log linear trend (see Altamirano \& Mendez 2015 for an analysis of the high-energy hard lags in GX 339-4)}. Again, we compared $\Delta\tau_{soft}$ with the disc-to-power law flux ratios and the 3-30 keV Eddington-scaled luminosities (Fig.~\ref{fig:lowhighf}, right panel).

The data clearly show that the amplitude of the low-frequency disc-leading feature increases as the source rises in luminosity through the hard state, and the disc-to-power law fraction increases. This same trend is observed also in the 0.2-0.8 Hz interval, though the increase of $\Delta\tau_{soft}$ with luminosity and disc relative flux is lower by a factor $\sim$1.5 (possibly as a consequence of the fact that these frequencies include contribution from higher-frequency variability components, see Sect. \ref{sect:lag-freq} and DM15, thus diluting the intrinsic lags associated with the low-frequency variability components).

On the other hand we find that the absolute amplitude of the reverberation lag decreases as the source rises in luminosity through the hard state, and the disc contribution increases. It is worth noting that the simple parametrization we adopted to describe the low-frequency disc-leading lag and the high-frequency reverberation lag is not a rigorous description of the intrinsic lag associated with each of these processes. This would require a detailed modeling of the lag-energy spectrum, which is beyond the aim of this paper. However, it is possible to show that, for small values of the phase lag (i.e. $\lsim 0.7$ rad, as measured for GX 339-4, see upper panels of Fig. \ref{fig:coherence}), the resulting lag is approximately equal to the weighted average of the phase lags associated with each process. The weights depend on the amplitude of variability of the different processes, and can be estimated from the covariance spectra (see Uttley et al. 2014; DM15). Thus, we computed the covariance spectra in the frequency intervals where the reverberation lag is observed, and fit them using the same model described in Sect. \ref{sect:rev}. We estimated the amplitude of variability of the reverberating component (i.e. the disc component) relative to that producing the hard lags log-linear trend (i.e. the power law component), and used these values to obtain an estimate of the intrinsic reverberation lag. We find slightly smaller values (respectively by $\sim$46 percent and $\sim$33 percent) of the reverberation lag during the high and medium luminosity observations than those shown in Fig.~\ref{fig:lowhighf} (right panel), while the value obtained for the low luminosity observation is consistent, within the errors, with that derived from our simple parametrization. However, even accounting for variations of the relative variability amplitude of the reverberation component with respect to the component producing the hard lags, the decreasing trend shown in Fig.~\ref{fig:lowhighf} (right panel) is still observed. This result points to a decrease of the intrinsic reverberation lag amplitude as a function of luminosity.

As pointed out in Sect. \ref{sect:lag-freq}, the reverberation lag maps relatively short distances, thus its amplitude is intrinsically small. For this reason, it is expected to dominate the lag-energy spectrum at relatively high frequencies, as indeed observed here. However, at low frequencies it should combine with the hard lags characterizing the soft band, thus slightly deviating (towards smaller absolute amplitudes) the measured lag from the intrinsic value. The magnitude of this shift depends on the amplitude of the reverberation lag. 
Since the absolute amplitude of the reverberation lag is observed to decrease as a function of luminosity (Fig.~\ref{fig:lowhighf}, right panel), the shift is expected to be smaller at higher luminosities. 
In fact, the observed increase with luminosity of the amplitude of the low-frequency disc-leading feature (Fig.~\ref{fig:lowhighf}, left panel) agrees with this picture and can be ascribed to the decreasing amplitude of the reverberation lag as a function of luminosity.

\begin{figure}
\centering
\includegraphics[width=9.cm]{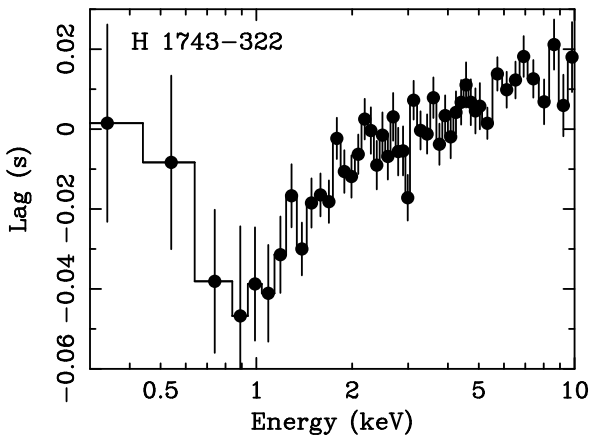} 
\caption{The lag-energy spectrum of H 1743-322 in the frequency interval 0.1-1 Hz.}
\label{fig:lagH1743}
\end{figure}

\section{Expanding the study of reverberation lags to the entire sample of BHXRBs}
\label{sect:other}

Following the same procedures described in Sects. \ref{sec:data}-\ref{sect:lag-energy} we carried out a systematic analysis of all the archived XMM-Newton observations (as of August 2014) of a sample of nine more BHXRBs. Our aim was to expand the study of reverberation lags to a larger number of sources and to verify whether the behaviours observed in GX 339-4 are common to all BHXRBs.
However, it turned out that the observations of GX 339-4 already presented in this paper are currently by far the best available for this kind of analysis. 
We found signatures of a reverberation lag only in one further observation, of H 1743-322 (Fig. \ref{fig:lagH1743}). This is due to the fact that most of the observations are characterized by low levels of variability S/N ratio and/or short exposures, thus hampering the detection of reverberation lags. Thus it remains possible that reverberation lags are a universal features of BHXRB X-ray emission, and we simply lack the data at present to confirm this. 

Results for each source can be summarized as follows:
\begin{itemize}
\item \emph{4U 1630-47:} the available XMM-Newton observations of this source (obsIDs: 0670671301, 0670671501, 0670672901-0670673201) are characterized by exposure times ranging between $\sim20-70$ ks. However, the fractional variability is only a few percent at all frequencies, so that the uncertainty on the lag is large. This precludes us the possibility to test for the presence of a reverberation lag. 
\item \emph{4U 1957+115:} as of August 2014 only one observation (obsID: 0206320101) was publicly available in the XMM-Newton archive (about 30 ks of effective exposure), which caught the source in a soft state (e.g. Nowak et al. 2008). The measured intrinsic variability is low at all frequencies and the error on the lags too large to obtain good constraints.
\item \emph{Cyg X-1:} of the 16 available observations, seven (obsIDs: 0202400101, 0202400501-0202401001) are either too short ($<$10 ks, thus giving very large error bars) or do not have EPIC pn data; four (obsIDs: 0202760201-0202760501) are not suitable for the analysis since they are taken in ``modified timing mode'', i.e. with the low-energy threshold set at $\sim$3 keV, thus not covering the disc emission (Duro et al. 2011);  observations 0202401101 and 0202401201 have very low fractional variability (a few percent) at all frequencies, thus the uncertainties on the lag measurements are large. Observations 0500880201, 0605610401 and 0610000401 have higher fractional variability ($\sim 10-20$ percent) but short effective exposures (i.e. $\sim$ 15 -24 ks). At low-frequencies ($\lsim2$Hz) the usual log-linear hard lag and the corresponding low-energy disc-leading feature are detected. A low-frequency (0.125-0.5 Hz) disc-leading feature  was first detected by Uttley et al. (2011) in the 0605610401 observation. However at high frequencies the lag errors are too large to allow us to test for the presence of a reverberation lag.
\item \emph{GRO J1655-40:} all the available observations of this source are either in soft (obsIDs 0112921301-0112921601, Sala et al. 2007a, and obsIDs 0155762501-0155762601, D\'iaz Trigo et al. 2007) or in quiescent states (obsIDs 0112460201, Reynolds et al. 2014, obsIDs 0400890301, and obsIDs 0400890201, Pszota et al. 2008). The fractional variability and/or the count rate are low thus increasing the uncertainties on the lag measurement.
\item \emph{GRS 1758-258:} there are three archived observations of this source (obsIDs: 0112971301, 0136140201, 0144630201), all with very short exposures ($\sim$20-8 ks), of which only one has high enough fractional variability, thus the uncertainties on the lags are large.
\item \emph{Swift J1753.5-0127:} The low-frequency hard lag is always detected in all the archived XMM-Newton observations (obs IDs: 0311590901, 0605610301, 0691740201, and 0694930501) of the source (in agreement with Uttley et al. 2011 and Cassatella, Uttley \& Maccarone 2012). However, given the relatively short exposures (i.e. $\sim$20-40 ks), the lag errors are too large to test for the presence of a reverberation lag at high-frequencies ($>$1 Hz). 
\item \emph{XTE J1650-500:} the source is in quiescence during the only observation available (obs ID 0206640101, Homan et al. 2006).
\item \emph{XTE J1817-330:} the source is in a disc-dominated, low-variability state during the only observation available (obs ID 0311590501, Sala et al. 2007b), thus the fractional variability is low.
\item \emph{H 1743-322:} there are two archived observations of this source (obs IDs 0553950201 and 0554110201). During observation 0553950201 the source is in quiescence. The other observation is characterized by high levels of fractional variability ($\sim$15-20 percent), typical of the hard state.
A soft-band upturn, ascribable to a reverberation lag, has been detected in the frequency range 0.1-1 Hz (Fig. \ref{fig:lagH1743}). 
However, the source is highly absorbed in the soft band, and the observation is relatively short ($\sim$ 21 ks), thus the S/N of the lag spectrum is reduced. For this reason this detection is less secure than those of GX 339-4.
\end{itemize}
Table \ref{tab:non-det} lists all the analyzed Epic-pn observations during which a reverberation lag was not observed, schematically summarizing the motivation preventing a detection.

\section{Discussion}
\label{sec:discuss}

From the analysis of simultaneous XMM-Newton and RXTE observations, we infer a remarkable energy-dependence of the X-ray lags in GX 339-4 (see Figs. \ref{fig:coherence} and \ref{fig:lagE}).
At high-energies ($E\gsim$2 keV) we find hard X-ray lags i.e. in the sense that higher energy photons lag behind lower energy photons, with a log-linear dependence on energy. This is in agreement with previous studies of BHXRBs (e.g. Miyamoto et al. 1988; Nowak et al. 1999; Pottschmidt et al. 2000). It is clear that high-energy, hard-lags are always observed where the power law dominates the spectrum, thus suggesting they are tied to this component. Their properties favor models of propagation of mass accretion rate fluctuations in the accretion flow (Lyubarskii 1997; Kotov, Churazov \& Gilfanov 2001; Ar\'evalo \& Uttley 2006, Ingram \& van der Klis 2013), as opposed to the initially proposed coronal Comptonisation models (e.g. Kazanas \& Hua 1999). In particular, their large amplitudes -- e.g. see Fig.~\ref{fig:coherence} -- would imply an unfeasibly large Comptonisation region (e.g. Nowak et al. 1999). According to propagation models, perturbations arising in the accretion flow at different radii propagate inwards and modulate the emission from the X-ray emitting regions.
Thus, one possibility to explain the hard lags is that the X-ray source is radially extended and has an energy-dependent emissivity profile. An emissivity profile that becomes more centrally concentrated as the energy increases, will lead to hard lags and a log-linear energy-dependence, as perturbations propagate from the outer to the inner radii (Kotov, Churazov \& Gilfanov 2001; Ar\'evalo \& Uttley 2006). Another possibility is that the corona is centrally concentrated, and the hard lags are due to variations of the power law photon index, as a consequence of variations of the seed photon luminosity modulated by the inward propagating perturbations (as suggested in Uttley et al. 2014).

At soft X-ray energies ($E\lsim$2 keV), in \emph{all} the observations the lags deviate significantly from a simple extrapolation of the trends observed at higher energies, as observed both from the lag-frequency, Sect. \ref{sect:lag-freq}, and the lag-energy, Sect. \ref{sect:lag-energy}, spectra. These deviations depend on the frequency interval considered and are clearly linked to the emergence of the disc component. In the lag-energy spectra we observe a disc-leading feature at low-frequencies and a reverberation lag at high-frequencies (see also Fig. \ref{fig:scheme}).
The low-frequency disc-leading feature appears as a steepening of the lag-energy spectrum in the soft band, meaning that the long-term variations in the disc component lead those of the power law by fractions of a second. This seems to be common in BHXRBs during the hard state (see Uttley et al. 2011 for the first detections of this feature), and might be related to the time needed for the corona emission to respond to variability of disc seed photons. On the other hand the reverberation lag, produced by thermal reprocessing, was known to be present only in the medium-luminosity observation of GX 339-4 (Uttley et al. 2011). 
Our results increase the number of detections of a reverberation lag in GX 339-4, and reveal a trend with luminosity.  Moreover, we find that a reverberation lag is present also in H 1743-322 (Sect. \ref{sect:other}), suggesting that this could be an ubiquitous feature.\\
Though the way BHXRBs evolve during outbursts is not yet well-assessed, an increase of the relative contribution of the disc to the total X-ray flux is commonly detected in the form of an increase of both thermal  (e.g. Dunn et al. 2010) and reflection (e.g. Plant et al. 2014) spectral components, as the source moves to higher luminosities through the hard state and from the hard to the soft state. In line with these results, we observe the disc luminosity increasing as the source rises in luminosity through the hard state. Assuming our simple model (Sect. \ref{sect:rev}), we estimate a factor $\sim$35 increase of the disc bolometric luminosity (E$=$0.01-1000 keV) from the low- to the high-luminosity observation.
This behaviour is usually ascribed to variations of the geometry of the inner accretion flow, and is consistent with a picture where smaller radii progressively start to contribute to the disc emission.
Within this scenario, the detected X-ray lags should vary accordingly.

\subsection{Low-frequencies}
\label{sect:disc1} 

The low frequency time lag-energy spectra of GX 339-4 can be explained assuming there are three variability processes at play (see Fig.~\ref{fig:scheme}).
One process is related to the hard X-ray Comptonization component and produces hard lags (positive lags with respect to our reference band, e.g. as described in models of propagation of mass accretion rate fluctuations, Kotov, Churazov \& Gilfanov 2001, Ar\'evalo \& Uttley 2006), with a log-linear energy dependence (Sect. \ref{sec:discuss}). The other process is associated with the disc component (e.g. due to propagation of mass accretion rate fluctuations throughout the disc; Uttley et al. 2011), and is responsible for soft photons leading the low frequency variability in the X-ray band. 
Such a process should cause correlated variations of the seed photons flux from the outer to the inner radii, which then modulate the low frequency variations of the hard X-ray Comptonization component. 
The last process is due to thermal reprocessing of the hard X-ray photons in the disc, and is responsible for producing small-amplitude, reverberation lags (see Sect. \ref{sect:disc2}). The observed lag-energy spectrum is determined by the average time lag of the impulse responses (e.g. Uttley et al. 2014) associated with the three processes in each energy bin. Within this picture, the reverberation lag is responsible for the observed flattening of the lag-energy spectrum that characterizes the data below $\sim$1 keV at low frequencies (Fig.~\ref{fig:lagE}, upper panels). Moreover, since the intrinsic amplitude of reverberation lag decreases with luminosity (Fig.~\ref{fig:lowhighf}, right panel), it is also responsible for the observed decrease of the amplitude of the disc-leading feature as a function of luminosity.

\subsection{High-frequencies}
\label{sect:disc2} 

At high-frequencies, the disc-leading component is suppressed (see Sect. \ref{sect:lag-freq}), and soft lags (i.e. whereby soft photons lag behind hard photons) are observed in all the analysed observations of GX 339-4. These are signatures of reprocessing, which starts to dominate at these frequencies. They appear precisely at the energies where the disc emission contributes $\gsim$40 percent of the total flux (Sect. \ref{sect:rev}).\\ 
At these frequencies, most of the disc photons come from the inner regions of the flow. Here irradiation of the disc by hard X-ray photons becomes important. The major effect of increased irradiation is an increase of reprocessed photons into thermal emission.
As a consequence, \emph{reverberation} is expected: variable hard X-ray emission will drive correlated high-frequency variability in the reprocessed component, with a delay approximately equal to the light travel-time between the two emitting regions.\\
Within this scenario, a change in the geometry of the system as the outburst evolves would determine a variation of the illumination pattern of the optically thick disc, as well as of the light travel-time from the primary source to the reprocessing region. 
The associated reverberation lag has to directly map these changes.

\begin{figure}
\centering
\includegraphics[width=9cm]{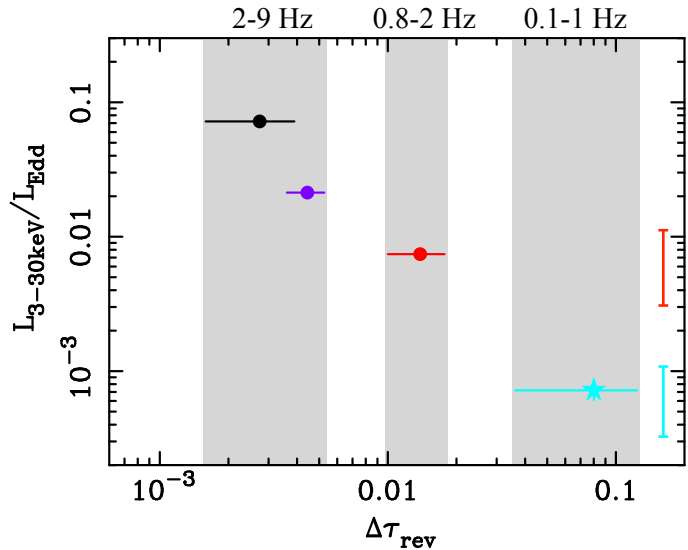}
\caption{The amplitude of the reverberation lag in GX 339-4 (black, blue, and red dots) and H 1743-322 (light blue star) as a function of luminosity (in units of Eddington luminosity). The shaded areas refer to the frequency intervals within which the reverberation lags have been detected, as listed on the top of the plot. 
The error bars on the right side of the plot are estimated taking into account the uncertainties on the distance and BH mass of the two sources, and show the relative difference in luminosity between H 1743-322 and the lowest-luminosity observation of GX 339-4.
}
\label{fig:highf_L}
\end{figure}

The high-frequency data show luminosity-dependent variations of the reverberation lag amplitude (Sect.~\ref{sect:Ldepend} and Fig.~\ref{fig:lowhighf}, right panel). In particular, the amplitude decreases as the source luminosity increases during the hard state.
This behaviour is consistent with a picture whereby the path from the primary source to the reprocessing region decreases as the outburst evolves. 
The distance traced by the reverberation lag can be expressed in terms of the number of gravitational radii, to give a scale-invariant measure of the direct-to-reprocessing region distance. However, this estimate is affected by the uncertainty on the BH mass. In the case of GX 339-4, only secure lower limits exist, namely $M_{BH}\gsim$6$M_{\odot}$ (Hynes et al. 2003; Mu\~noz-Darias, Casares \& Mart\'inez-Pais 2008). Another relevant but unknown parameter is the inclination angle of the source. However, based on the lack of eclipses or dipping, the shape of the tracks on the hardness-intensity diagram (Mu\~noz-Darias et al. 2013 and  figure 1 in DM15) and the non-detection of accretion disc winds traced by high ionisation Fe K lines in GX 339-4 (Ponti et al. 2012b), we can infer that this system has a small inclination. 
Assuming the value of 8$M_{\odot}$ for the BH mass (Sect. \ref{sec:data}) and considering our lag measurements (and the corresponding uncertainties) at different luminosities, we obtain the range of distances of $\sim$35$\pm15 r_{g}$ to $175\pm 43 r_{g}$ (it is worth noting that these are order of magnitudes estimates, derived from the simple parametrization of the reverberation lag adopted in Sect.~\ref{sect:Ldepend}, and considering only light-travel time delays; a more accurate estimate of the intrinsic distance to the reprocessing region would require taking into account the caveats discussed in Sect. \ref{sect:Ldepend}, as well as a detailed modeling of the lag-energy spectrum, which is beyond the aim of this paper).
These values imply a relatively large distance between the primary X-ray source and the reprocessing region. This is expected in the case of a truncated disc or a large coronal height, though we note that a quasi-stationary corona powered by magnetic fields anchored to the disc is unlikely to have a height larger than $\sim 30$ r$_g$ (Fabian et al. 2014). In the assumption of a small coronal height and taking the estimated distances as an approximate measure of the reprocessing radius, we notice that the observed trend with luminosity is in agreement with that characterizing the inner radius parameter as estimated from the reflection component during more evolved states of the source (see figure 9 of Plant et al. 2014).
Thus, in order to explain the data, we need the optically thick disc to be truncated, or its inner radii to be not directly observable and/or strongly irradiated by the hard X-ray source. Hot flow models, where a hot inner accretion flow is surrounded by a truncated optically-thick cold accretion disc, are viable solutions for the first case (e.g. Done, Gierli\`nski \& Kubota 2007 and references therein). Alternatively, assuming the cold disc is not truncated in the hard state, then an optically thick (i.e. with optical depth at least $\sim$1) intervening medium (e.g. a non-homogeneous corona) is needed to destroy coherence between the directly observed primary emission and the fraction that irradiates the inner disc regions, preventing the detection of reverberation lags from the smallest radii (to this end note that estimates of the average optical depth of the corona as measured from hard state spectra give values of $\sim 0.6-1$, higher than in softer states, e.g. Done 2010). In both cases, the variations of the reverberation lag intrinsic amplitude with the outburst evolution imply that the reprocessing region gradually approaches the ISCO, i.e. either the truncated-disc radius decreases or the optically thick corona recedes/becomes gradually optically thinner. Alternatively, the observed variations of reverberation lag amplitude might be driven by variations of the height of a centrally concentrated corona as a function of luminosity (e.g. Fabian et al. 2014).\\
Finally, following the same parametrization procedure as in GX 339-4 (Sect. \ref{sect:rev}), we estimated the amplitude of the reverberation lag in H 1743-322 and compared it to the values obtained from the analysis of GX 339-4. In Fig.~\ref{fig:highf_L} we plot the results as a function of 3-30 keV Eddington scaled luminosity (the Eddington luminosity has been computed using the estimate of 8.5$\pm$0.8 kpc for the distance, Steiner, McClintock, \& Reid 2012, and assuming a standard mass range of $5-15$M$_{\odot}$)\footnote{Note that in Fig.~\ref{fig:highf_L}, when comparing the Eddington scaled luminosities of the different observations of GX 339-4, the relevant error is the error on the flux measurements (i.e. $\sim0.1-0.3$ percent). However, when the luminosity of GX 339-4 is compared with the luminosity of H 1743-322, the uncertainty on the distance and BH mass estimates of the two sources must be taken into account. The corresponding error bars are reported in Fig.~\ref{fig:highf_L} to show the relative difference in luminosity between H 1743-322 and the lowest-luminosity observation of GX 339-4.}.
We find that the lag in H 1743-322 is perfectly in agreement with the values and trend observed in GX339-4, indicating that this is not a peculiar property of one source, but a behaviour likely common to BHXRBs, although the lack of suitable data for other systems prevents definitive conclusions on this point.

\section{Summary \& Conclusion}
The aim of the study presented in this paper is to search for and monitor reverberation lags in BHXRBs as a function of the accretion state. However, most of the observations turned out to be unsuitable for this kind of analysis. Only GX 339-4 could be studied in detail, with a total of four data sets which sampled the source at three different levels of luminosity during the hard state. 
This has allowed us to expand the analysis presented in Uttley et al. (2011), and to follow the evolution of the lags as a function of X-ray luminosity throughout the hard state.
Our main conclusions can be summarized as follows:
\begin{itemize}
\item The X-ray lags of GX 339-4 display a significant contribution from the disc component, superimposed on the hard continuum lags. The disc component leads the hard X-ray variability at low-frequencies, while it lags behind (reverberation) the hard X-ray Comptonization component at high-frequencies (in agreement with Uttley et al. 2011), as a consequence of thermal reprocessing in the disc. This behaviour, irrespective of luminosity, is a characteristic feature of the source during the hard state.
\item At low frequencies/long time scales, the reverberation lag combines with the hard lags associated with the disc and the power law component (and possibly due to propagation of mass accretion rate fluctuations), and is responsible for the flattening of the lag-energy spectra at E$\lsim$1 keV.
\item The reverberation lag is clearly detected, at high frequencies/short time scales, in all the observations of GX 339-4. The lag intrinsic amplitude decreases with the increase of disc-fraction as the luminosity of the source increases. These variations of reverberation lag amplitude affect also the low-frequency lag-energy behaviour, causing the low-frequency disc-leading feature in the soft X-ray band to vary with luminosity accordingly.
\item Our findings imply that the reprocessing region gradually moves to smaller radii, and approaches the ISCO as the source rises in luminosity through the hard state. Alternatively, a corona whose height decreases with luminosity might explain the observed behaviour.
\item The reverberation lag detected in H 1743-322 is consistent with the trend of lag amplitude decreasing with luminosity observed in GX 339-4.

\end{itemize}

Finally we note that the described behaviour of X-ray lags should represent a general property of transient BHXRBs. However, the currently available observations do not allow us to definitely test this hypothesis and derive more general conclusions. Specifically designed, XMM-Newton monitoring campaigns of BHXRB outbursts would be needed to accomplish this goal.

\acknowledgments
The authors thank the anonymous referee for helpful comments which significantly improved the paper. The authors also thank Phil Uttley for useful and constructive discussion. This work is based on observations obtained with XMM-{\it Newton}, an ESA science mission with instruments and contributions directly funded by ESA Member States and NASA. GP acknowledges support via the Bundesministerium f\"ur Wirtschaft und Technologie/Deutsches Zentrum f\"ur Luft und Raumfahrt (BMWI/DLR, FKZ 50 OR 1408) and the Max Planck Society. TMD acknowledges support by the Spanish Ministerio de Econom\'ia y competitividad (MINECO) under grant AYA2013-42627.

{\it Facilities:} \facility{XMM (Epic-pn)}, \facility{RXTE (PCA)}.

\clearpage
\begin{turnpage}
\begin{deluxetable*}{ccccccccccccc}
\tabletypesize{\tiny}
%\rotate
\tablecaption{Reverberation lags: non-detections.\label{tab:non-det}}
\tablewidth{0pt}
\tablehead{
\colhead{Source} & \colhead{ObsID} & \colhead{low} & \colhead{low} & \colhead{short} & \colhead{hard}  & &
\colhead{Source} & \colhead{ObsID} & \colhead{low} & \colhead{low} & \colhead{short} & \colhead{hard} \\
\colhead{} & \colhead{} & \colhead{fract-var} & \colhead{count rate} & \colhead{exp} & \colhead{lag}  & &
\colhead{} & \colhead{} & \colhead{fract-var} & \colhead{count rate} & \colhead{exp} & \colhead{lag} }
\startdata
GX 339-4   	 &    0093562701 	&  \checkmark & 		          & 			   & 		  &			 &		Cyg X-1             & 0605610401 &		 	     &                    &\checkmark & \checkmark \\
GX 339-4   	 &    0111360201	&                	&\checkmark & 			   & 		  &			 &		Cyg X-1              & 0610000401 &		 	&                    &\checkmark & \checkmark \\
GX 339-4   	 &    0111360501	&                	&\checkmark & 			   & 		  &			 &	GRO J1655-40         &  0112460201 &		 	&\checkmark &		 & 		 \\
GX 339-4   	 &    0148220201	&  \checkmark & 		          & 			   & 		  &			 &	GRO J1655-40         & 0112921301  &\checkmark&                    &		 & 		 \\
GX 339-4   	 &    0148220301	&  \checkmark & 	           	    & 			   & 		  &			 &	GRO J1655-40         & 0112921401   &\checkmark&                    &		 & 		 \\
GX 339-4   	 &    0156760101	& \checkmark  & 	       	    & 			   & 		  &			 &	GRO J1655-40         & 0112921501  &\checkmark &                    &		 & 		 \\
GX 339-4   	 &    0410581201	& \checkmark  & 	      	    & 			   & 		  &			 &	GRO J1655-40          & 0112921601  &\checkmark &                    &		 & 		 \\
GX 339-4   	 &    0410581301	&  \checkmark & 		          & 			   & 		  &			 &	GRO J1655-40          & 0155762501  & \checkmark&                    &		 & 		 \\
GX 339-4   	 &    0410581701	&  \checkmark & 		          & 			   & 		  &			 &	GRO J1655-40          & 0155762601  & \checkmark&                    &		 & 		 \\
4U 1630-47	 & 0670671301 	& \checkmark 	& 		          & 			   & 		  &			 &  GRO J1655-40         &  0400890201  &		         & \checkmark&		 & 		 \\
4U 1630-47	 & 0670671501	& \checkmark  & 		          & 			   & 		  &			 &  GRO J1655-40          & 0400890301  &                   &\checkmark &		 & 		 \\
4U 1630-47	& 0670672901		& \checkmark  & 		          & 			   & 		  &			 &  GRS 1758-258          & 0112971301  &                     &                    &\checkmark &		 \\
4U 1630-47	& 0670673001 	& \checkmark & 		          & 			   & 		  &			 &  GRS 1758-258         & 0136140201  &                    &                    &\checkmark  & 		 \\
4U 1630-47	& 0670673101	     & \checkmark   & 		          & 			   & 		  &			 &	GRS 1758-258         & 0144630201 & 			   &			         & \checkmark &		 \\
4U 1630-47	& 0670673201		& \checkmark  & 		          & 			   & 		  &			 &   Swift J1753.5-0127 & 0311590901  &		        & 			         & \checkmark &\checkmark\\
4U 1957+115 & 0206320101       & \checkmark & 		          & 			   & 		  &			 &	Swift J1753.5-0127  & 0605610301  & 				&			    & \checkmark &  \checkmark \\
  Cyg X-1	    & 0202400501        &  			    & 		          & \checkmark &		        & 	       	&   Swift J1753.5-0127  & 0691740201  &                      &                   & \checkmark &  \checkmark\\
  Cyg X-1     	&  0202400601      &  			     & 		         & \checkmark &		        & 	       	&   Swift J1753.5-0127  &  0694930501  &                     &                    	& \checkmark &  \checkmark \\
Cyg X-1		&  0202401101    &\checkmark   &                   &                          &              &              &   XTE J1650-500        & 0206640101   &                      & \checkmark  &                       &          \\
Cyg X-1		& 0202401201   & \checkmark   & 		        &                       & 	        &	            &   XTE J1817-330       &  0311590501   & \checkmark   &                     &                       &           \\ 
Cyg X-1		&  0500880201   &  			    &	               & \checkmark & \checkmark&           &  H 1743-322      & 0553950201     &                      & \checkmark &                       &           \\    
\enddata
\tablecomments{All the analyzed Epic-pn XMM-Newton observations during which a reverberation lag was not detected.
For each source and observation ID, the table summarizes the causes that prevented the detection of a reverberation lag (i.e. low fractional variability, low count rate, or short exposure). 
The table also reports in which of these observations the low-frequency disc-leading feature and/or log linear hard-lag have been detected.}
\end{deluxetable*}
\end{turnpage}
\clearpage
\global\pdfpageattr\expandafter{\the\pdfpageattr/Rotate 90}

\end{document}